\documentstyle[11pt,aaspp4]{article}

\begin{document}

\def\mathnew{\mathsurround=0pt}
\def\simov#1#2{\lower .5pt\vbox{\baselineskip0pt \lineskip-.5pt
        \ialign{$\mathnew#1\hfil##\hfil$\crcr#2\crcr\sim\crcr}}}
\def\simg{\mathrel{\mathpalette\simov >}}
\def\siml{\mathrel{\mathpalette\simov <}}
\def\Mesz{M\'esz\'aros~}
\def\beq{\begin{equation}}
\def\enq{\end{equation}}
\def\bea{\begin{eqnarray}}
\def\ena{\end{eqnarray}}
\def\bec{\begin{center}}
\def\enc{\end{center}}
\def\hl{\hline}
\def\tl{\tableline}
\def\half{ 1/2 }
\def\lbr{\linebreak\noindent}
\def\L51{L_{51}}
\def\r13{r_{13}}
\def\et2{\eta_2}
\def\Gam2{\Gamma_2}
\def\th2{\theta^2}
\def\T7{T_7}
\def\barnb{ {\bar n_b} }
\def\barSigb{{\bar \Sigma_b}}
\def\Gamb{\Gamma_b}
\def\Gamb2{\Gamma_{b2}}
\def\Gam2{\Gamma_2}
\def\calN{{\cal N}}
\def\HI{\hbox{HI}}
\def\HeI{\hbox{HeI}}
\def\HeII{\hbox{HeII}}
\def\FeXXVI{\hbox{FeXXVI}}
\def\varep{\varepsilon}

\received{3/27/98 }
\accepted{6/3/98 }
\slugcomment{ApJL, acc. 6/3/98}
%\journalid{XXX}{XXX 1997}
%\articleid{11}{14}
\lefthead{ \Mesz \& Rees }
\righthead{ Spectral Features from Ultrarelativistic Ions in GRB}
\tighten

\title{ SPECTRAL FEATURES FROM ULTRARELATIVISTIC IONS \\
IN GAMMA-RAY BURSTS ? }

\author{P. \Mesz } %\altaffilmark{1} 
\affil{Dpt. of Astronomy \& Astrophysics, Pennsylvania State
University, University Park, PA 16803}

\and

\author{M.J. Rees} %\altaffilmark{2} }
\affil{Institute of Astronomy, Cambridge University, Madingley Road,
Cambridge CB3 0HA, U.K.}

%\bec Draft :~~{\today} \enc
%\newpage

\begin{abstract}

Gamma ray burst outflows may entrain small blobs or filaments of dense, 
highly ionized metal rich material. Such inhomogeneities, accelerated by 
the flow to Lorentz factors in the range 10-100, could have a significant
coverage factor, and give rise to broad features, especially due to Fe 
K-edges, which influence the spectrum below the MeV range, leading to 
a progressively decreasing hardness ratio.

\end{abstract}


%\newpage

\section{Introduction}

Gamma-ray bursts (GRB) have been detected in the past year at X-ray, as well
as optical and radio frequencies (e.g. Costa, et al. 1997, Sahu, et al.
1997, Frail et al. 1997; recent results are summarized in Meegan, Preece 
\& Koshut, 1997). A cosmological origin is indicated by the measurements of 
redshifts in at least two objects (Metzger, et al. 1997, Kulkarni, et al. 
1998). The radiation is generally interpreted in terms of nonthermal 
continuum emission from shocks in a relativistic fireball outflow, both 
in the early high energy emission (Rees \& \Mesz, 1992; \Mesz \& Rees, 
1993; Piran, Shemi \& Narayan, 1993; Katz, 1994; Rees \& \Mesz, 1994; Sari 
\& Piran, 1995; Papathanassiou \& \Mesz, 1996; Panaitescu, Wen, Laguna \& 
\Mesz, 1997) and in the subsequent afterglows at longer wavelenghts (\Mesz 
\& Rees, 1997a; Vietri, 1997; Waxman, 1997; Wijers, Rees \& \Mesz, 1997).  
While the outflow is typically
assumed to be chemically homogeneous and smooth on average (except for 
instabilities and shocks), it could have a substantial component of blobs 
of denser material (e.g. from the small mass fraction near the surface of a 
disrupted neutron star torus) which are entrained by the average outflow, 
and coexist with it in pressure equilibrium. This denser material would be 
richer in heavy elements, and could have significant spectral effects 
caused by absorption edges from metals such as Fe, with consequences for 
the early $\gamma$-ray and X-ray emission from GRB (the related effects 
in GRB afterglows will be discussed elsewhere). In what follows we 
investigate the physical conditions in such blobs, and calculate the effects 
they have on the observed spectrum associated with internal shocks in GRB. 

\section{Baryonic Outflow and Dense Blob Entrainment }

In a fireball outflow arising from the disruption of a compact binary
or the collapse of a fast rotating stellar core, internal shocks and
nonthermal radiation leading to $\gamma$-ray emission arise at radii
$r_{sh} = c t_v \eta^2 = 3\times 10^{14} t_v \et2^2$ cm, where $t_v \simg
10^{-3}$ s is the variability timescale and $\eta=10^2 \et2$ is the terminal
coasting bulk Lorentz factor, determined by the baryonic loading of the
outflow.  We do not know to what extent the outflow is
beamed, but for the present discussion we suppose it  is  confined inside
channels of solid angle $\theta^2$. For a total luminosity $L=10^{51} \L51$  and
mass outflow rate $\dot M = L /( c^2 \eta )$ lasting for a time $t_w \simg t_v$,
the mean comoving  density  of nuclei  in the smooth outflow is
\beq
n_o = (L/4\pi \theta^2 r^2 \eta^2 A m_p c^3)= 3\times 10^{12}L_{51}\theta^{-2}
r_{13}^{-2} A_o^{-1} \eta_2^{-2} ~\hbox{cm}^{-3}~,
\label{eq:no}
\enq
where $r = 10^{13} \r13$ and $A_o$ is the mean particle atomic weight.
The total baryonic mass per unit logarithmic radius is $ M_o = 4\pi \theta^2
r^3 \eta^{-1} n_o A_o m_p = 4\times 10^{26} L_{51} \eta_2^{-3} r_{13}^2 A_o$ g,
and the corresponding smoothed- out  column density  of nuclei is
\beq
\Sigma_o = 3\times 10^{23} \L51 \r13^{-1}\theta^{-1} A_o^{-1} \et2^{-3}~~
\hbox{cm}^{-2}~.
\label{eq:Sigmao}
\enq

 The outflow can also carry magnetic fields whose comoving energy
density in the frame moving with $\eta$, expressed as a fraction $\xi_B$
of the total energy density, gives $B= 3 \times 10^5  L_{51}^{1/2} \xi_B^{1/2}
\theta^{-1} r_{13}^{-1} \eta_{2}^{-1}$. If the outflow is magnetically-driven
from the central object, then $\xi_B$ would be not much less than unity. An
important consequece of such strong fields is that the gyroradii are small. This
means that the flow can be treated as fluid-like. Moreover, conductivity and
diffusion are severely inhibited, at least across the field, so that blobs or
filaments of cooler and denser material  could exist, in pressure balance with
their surroundings. (This possibility has been discussed in other contexts by
Celotti et al., 1998).
  
In addition to a smooth distribution of baryons, dense blobs of (possibly 
Fe-enriched) matter may be able to survive and be entrained in the flow. 
A small blob moving with bulk Lorentz factor $\Gamma_b$  (possibly less than 
$\eta$) whose gas temperature was of order of the comoving photon temperature 
$T \sim 10^7 \T7 \Gamma_{b2}^{-1}$ K (or $\sim$ 100 keV in the observer frame) 
could have a particle density (measured in its own comoving frame) of up to
\beq
n_b \simeq  2\times 10^{18} L_{51} 
\theta^{-2} r_{13}^{-2} T_7^{-1} \Gamma_{b2}^{-2} ~\hbox{cm}^{-3}~,
\label{eq:nb}
\enq
This maximum density would be reached if its internal pressure  balanced  the
total external (magnetic and particle) pressure. If the blobs were composed of
iron-rich material from neutron-star debris, the density of nuclei would be
lower than $n_b$ by a factor  $1/Z_b$,  the average charge of the ions.
Such blobs are much denser than the corresponding ``background"  baryon
density given in equation (\ref{eq:no}). We return in \S 4 to discuss  the
internal thermal balance, and to show that they could indeed remain with $T_7
\siml 1$. However, we first consider the geometry and dynamics of such blobs.

Suppose the  blobs have a volume filling factor $f_v = {\bar n_b}/n_b$. 
This is of course likely to be a very small number. However, if the  blobs are 
individually very small, the surface covering factor $f_s$ can nonetheless be 
substantial. If the blobs were spheres of characteristic radius $r_b$ then  
$r_b n_b f_s$  would equal the smoothed-out column density over one comoving 
length scale $c t_{exp}=r/\Gamma_b$ in the frame of the blobs, $\barSigb =
{\bar n_b} (r/\Gamma_b)$. We 
obtain $r_b=(r/\Gamma_b)(\barnb/n_b)f_s^{-1}=(r/\Gamma_b)f_v f_s^{-1}$. If one 
sets the smoothed-out density of blobs moving at $\Gamma_b$, as seen in the
flow frame $\eta$, equal to a fraction $\alpha$ of the average flow comoving 
particle density, $\barnb=\alpha n_o \eta \Gamma_b^{-1}$, the volume filling
factor is $f_v=1.5 10^{-6} \alpha \et2^{-1}\Gamb2 A_o^{-1}\T7$. The blob size
$r_b=\barSigb/(n_b f_s)$ is given by
\beq
r_b= \barSigb /(n_b f_s) = 1.5\times 10^5\alpha\r13 \et2^{-1}
A_o^{-1} \T7 f_s^{-1}~\hbox{cm}~,
\label{eq:rb}
\enq
while the column density through a single blob is just $\Sigma_b=r_b n_b=
\barSigb f_s^{-1}$, and the average smoothed out column density from blobs is 
$\barSigb= \alpha \Sigma_o (\eta/\Gamma_b)^2$ in the $\Gamma_b$ frame. In order 
to have a surface coverage factor $f_s >1$, there is an upper limit $r_b < 
\barSigb/n_b$ on the blob sizes. Realistically, the blobs are likely to be 
streaks or filaments elongated along the magnetic field direction, the field 
itself being predominantly perpendicular to the radial direction.
The above formulae carry over provided we identify $r_b$ with the smallest
dimension: this is likely to be the dimension transverse to the field, and can
readily be small enough to permit a large covering factor, while nonetheless
being  large enough (compared to the gyroradius) to ensure a fluid-like
behavior.

\section{Blob Velocities}

The blobs, even if consisting of gas entrained from a slower moving 
environment, will tend to be accelerated  by the  mean MHD jet outflow.
This flow starts at some lower radius $r_l =10^6 r_{l,6}$ cm, and reaches
its saturation bulk Lorentz factor $\eta = L/({\dot M} c^2)= 10^2 \et2$ well
before internal shocks reconvert a significant fraction of the bulk kinetic
energy  into radiation at radii $r_{sh} \sim 10^{13}\r13$ cm; still further out, 
there may be a deceleration shock where the ejecta encounter the external 
medium.  Blobs entrained into the flow near $r_l$, or from the boundary of the 
channel at larger radii,  are accelerated by the flow; at or above the shock 
radius Compton scattering of the intense photon flux is of comparable 
importance for the dynamics. 

The comoving radiation energy density in the flow is $u_\gamma =L/(4\pi 
\theta^2 r^2 c \eta^2) = 3\times 10^9 \L51 \r13^{-2} \theta^{-2} \eta_2^{-2} ~
\hbox{erg ~cm}^{-3}$.  The radiation pressure would accelerate any 
optically-thin blob into a frame  in which the net Compton drag were zero, 
on a  timescale $t_{dr}= A m_p c^2/(\sigma_T c u_\gamma )= 4\times 10^1 
\L51^{-1} \r13^2 \theta^2 \eta_2^2 A ~\hbox{s}~$.
This  timescale (calculated taking account  of the inertia of the ions, which
are, on the macroscopic level, constrained to move with the leptons)
is shorter than the comoving expansion (dynamic) time of the flow
$t_{ex}=r(c\eta)^{-1}=3~ \r13 \eta_2^{-1} ~\hbox{s}$ for radii $r_{13} \siml 
0.75\times 10^{-1} \L51 \eta_2^{-3}\theta^{-2} A_o^{-1}$.
For an optically thin blob released at some radius $r_o$ the terminal Lorentz
factor achievable is (Phinney 1987) $\Gamma_{b,max}= (L / L_{Ed})^{1/3} \sim
2\times 10^4 \L51^{1/3}$. For an optically thick blob, the effective
acceleration is lowered by a factor ${\calN_o}^{-1} = (\Sigma_{bo}/
1.5\times 10^{24} \hbox{cm}^{-2})^{-1}$, so
$\Gamma_{b,max} = (L/L_{Ed})^{1/3} {\calN_o}^{-1/3} = 2\times 10^4
\L51^{1/3} {\calN_o}^{-1/3}$.
%\label{eq:Gammabmx}
Although the above expressions refer to radiation-pressure acceleration,
similar considerations apply to acceleration by the ram pressure and Poynting
flux of  the smooth relativistic outflow. If $L$ is defined as the total
energy flux, the results are identical provided that $\calN_o$ exceeds
$1.5\times 10^{24}$.  

%CH
A blob immersed in a hydromagnetic flow carrying a flux $L$
behaves in a similar way. If its column density is sufficiently
low, its motion adjusts to the same Lorentz factor as the
surrounding flow. The condition for this to happen is that
\beq 
{\calN_o} < (L/L_{Ed}) (r/r_o)^{-1} \Gamma_b^{-3}~.
\label{eq:calN}
\enq
Note that the dependence on $\Gamma_b$ arises because, if the blob
moved with a slightly different speed from the mean flow, the
drag force on it (in the comoving frame) scales as $r^{-2}\Gamma_b^{-2}$
and the time available, at a given $r$, scales as $r \Gamma_b^{-1}$.

A blob for which $\calN_o$ is low enough to satisfy the condition 
(\ref{eq:calN}) at the radius where the velocity of the mean outflow saturates 
will coast stably outwards in pressure balance with its surroundings.
Blobs with higher $\calN_o$, for which (\ref{eq:calN}) is not satisfied,
would be accelerated by the ram pressure associated with
the energy flux $L$, but would not attain the same Lorentz factor
as their surroundings. The thickness of such blobs would adjust
to be equal to the scale height corresponding to the acceleration,
which would be proportional to the blob temperature $T$, and also 
proportional to $\calN_o$ times $\Gamma_b^2$. 
 
Thus we expect that the flow, out at radii $\sim 10^{13}$ cm, would
contain small blobs with Lorentz factor of order $\eta$, and
also larger blobs with lower Lorentz factors. As we discuss
later, this slower-moving material could have an important
effect on the time-evolution of the spectra of gamma ray bursts.
 
The proportions of slow-moving and fast-moving blobs would
depend on the uncertain details of how the initial entrainment
occurs, and also on the effects of instabilities during the
outflow. Blobs small enough to satisfy the condition (\ref{eq:calN}), which 
in effect constitute a ``mist" of clouds or filaments embedded in the
flow (preserved by strong magnetic fields against diffusion
effects), are not subject to any obvious dynamical instability.
Larger blobs, on the other hand, would seem in principle vulnerable
to both Rayleigh-Taylor and Kelvin-Helmholtz instabilities.
 
However, in a magnetically-dominated outflow, acceleration
of blobs could plausibly occur without triggering Rayleigh-
Taylor instability. The situation could be analogous to, for
instance, solar prominences, where magnetic stresses support
cool gas against gravity (c.f. the classic work of Kippenhahn
\& Schl\"uter, 1957, and many later variants). Kelvin-Helmholtz 
instabilities are more problematic: even though these tend to be 
suppressed by magnetic fields with a component along the flow direction
(e.g. Hardee, et al., 1992) or by fields in the blobs themselves, it is 
unclear to what degree they are, and it is unlikely that they can be 
eliminated completely.  What we are envisaging is a more
extreme version of what we know is going on in SS433 (where
a combination of mass flux and emissivity constraints forces
one to a model involving cool blobs with small volume-filling
factor accelerated to 10,000 times their internal sound speed).
 
The range of blob sizes (and blob Lorentz factors) at $10^{13}$ cm
will therefore depend on (a) the nature of the entrainment process:
(b) the extent to which slower (heavier) blobs are shredded by
Kelvin-Helmholtz instabilities; and (c) the possible countervailing
effect of coalescence. which can be important when the covering
factor is of order unity and a range of velocities is present.
We regard this as an open question, and turn now to consider the
thermal equilibriun within blobs, which depends primarily on
the radiation field and the pressure.
%CHE

\section{Temperature and Ionization State}

The ionization rate is expected to be extremely high in a GRB outflow,
but the blobs are so dense that the recombination rate is exceptionally 
high as well.  This has two important consequences. First, the
'ionization parameter', which  depends on the ratio of ionization and
recombination rates, and determines the equilibrium state of ionization in the
blobs, is not vastly different from what is familiar in some X-ray sources.
Second, because the recombination timescale is so short, each electron can
recombine (and be reionized)  during the outflow timescale, so the  blobs can
reprocess most of the photon flux from a burst, even though their total mass 
is low.

The ionization parameter $\Xi =L/ n_b r^2$ (e.g. Kallman \& McCray, 1982),
evaluated in the comoving frame, is $\Xi =L/ n_b r^2 \Gamma_b^2 = 5\times 10^2 
\theta^2 \T7 $.  For $\Xi \simg 10^3$, most Fe would be present as FeXXVI (i.e. 
H-like) or fully stripped; this would still be true if the material were so 
enriched in Fe that this is the dominant species.
Self-shielding would be inevitable if the total number of recombinations
became comparable with the number of ionizing photons available. The total
number of recombinations per second per unit logarithmic radius for a plasma 
with mean ionic charge $Z_b$ is ${\cal R}_r \sim \alpha n_i n_e V f_v$, where 
$V= 4\pi\theta^2 r^3\eta^{-1}$, $n_e \simeq n_b$ is electron density, 
$n_i= n_b /Z_b$ is ion density, and $\alpha \sim 2\times 10^{-11} Z^2 T^{-1/2}$ 
is the recombination coefficient for hydrogenic ions.
The total number of ionizations per second in the same volume will be
approximately equal to the number of ionizing photons injected per second
above the shock region, ${\cal R}_i \sim L /(\Gamma_b^2 h \nu_i)$, where, for
10 KeV photons in the comoving frame, $h\nu_i \sim 10^{-8}\varep_{10}$ erg. Thus
\bea
{\cal R}_r & \simeq & 4\times 10^{54} \alpha \L51^2 \xi_B\r13^{-1}\theta^{-2}
\et2^{-2}\Gamb2^{-3} A_o^{-1} \T7^{-3/2} Z_b ~~\hbox{s}^{-1}~; \\
{\cal R}_i & \simeq & 10^{55}\Gamb2^{-2} \varep_{10}^{-1}~~\hbox{s}^{-1}~.
\label{eq:ioniz}
\ena
If the blob parameters were such that ${\cal R}_r \simg {\cal R}_i$, the
optically thin assumption would not be self-consistent, and self-shielding
could be important.  Bound-free and bound-bound line cooling could then 
have an additional effect in determining the blob temperature. However, the 
blobs cannot cool below the black-body temperature of the comoving radiation 
field  $u_\gamma =3\times 10^9 \L51 \r13^{-2} \theta^{-2} \eta_2^{-2}~
\hbox{erg ~cm}^{-3}$, which is $T_{bb} \sim 10^6 \L51^{1/4}(\r13\theta 
\et2)^{-1/2}$ K; this suggests that H will always be almost completely
ionized by collisions. Note also that absorption by ions in the diffuse flow
is negligible, because for a given total mass the recombination rate in blobs 
is larger by the same ratio as the densities.

We have already shown that small blobs could contribute a covering factor
of order unity.  In conjunction with the above inference that the recombination
rate can be comparable with the total photon production rate, this tells us that
the blobs could 'reprocess' much of the radiation. The optically thin estimate 
% CH
(\ref{eq:ioniz}) and the above temperature estimates indicate that, 
independently of any self-shielding, substantial recombinations of highly 
ionized heavy elements such as Fe would be expected.  They can thereby create 
absorption features, the absorbed energy being re-emitted as (very broadened) 
lines.
%CHE

\section{Optical Depth and Spectral Widths}

Absorption edges are expected to form at energies corresponding to the 
K-$\alpha$ absorption of hydrogenic ions. 
The hydrogenic photoionization cross section is $\sigma_{th} \simeq 8 \times 
10^{-18} Z^{-2} \hbox{cm}^2$ at the threshold $h\nu_{th} \simeq 13.6 Z^2$
eV, decreasing above that as $(\nu/\nu_{th})^{-3}$. E.g., for FeXXVI the
threshold in the blob frame is at 9.28 KeV, and the cross section is 
$\sigma_{th} \sim 1.2\times 10^{-20}$ cm$^2$. Multiplying by the mean ion 
column density from blobs $\barSigb / Z_b =\alpha\Sigma_o Z_b^{-1}
(\et2/\Gamb2)^2$ (equation [\ref{eq:Sigmao}]), for hydrogenic ions the mean 
optical depth and the observer frame threshold energy are
\bea
\tau_{th} & \simeq & 1.4\times 10^{2}\alpha\L51\r13^{-1} \theta^{-2} A_o^{-1}
        \et2^{-1} \Gamma_{b2}^{-2} x_i (Z_b/26)^{-3}~~; \\
h\nu_{th}  & \simeq & 0.928 ~(Z_b/26)^2 \Gamb2~~\hbox{MeV}~,
\label{eq:nuedge}
\ena
where we normalized to Fe XXVI blobs, $x_i$ being the ionic abundance fraction
by number.  For Fe XXV the optical depth would be similar, modulo the 
ionization fraction, and the threshold is at $.883\Gam2$ MeV, while for HeII 
the optical depth could be larger, if $\Xi \siml 10^2$, and the edge would be 
at $0.544\Gamb2$ KeV. (An HI edge at $0.136\Gamb2$ KeV might just be possible 
if $\Xi \siml 50$ for cooler blobs at larger radii).
Bluewards of the absorption edges one would expect the flux to be blanketed
up to a comoving photon energy $\nu_{max}$ such that $\sigma_{th}
(\nu_{max}/\nu_{th})^{-3} (\barSigb /Z_b) =1$, where it gradually rejoins 
the continuum level. 

In addition to edges, K-$\alpha$ resonant features are also expected at energies
redwards of the edges, e.g. at comoving energies of 6.9 KeV for FeXXVI, or 
$0.69\Gam2$ MeV in the observer frame. The expected equivalent width in the 
damping wing dominated regime is $(W_\nu / \nu ) \simeq 0.15~(\alpha\L51
\r13^{-1}\theta^{-2} A^{-1} Z_b^{-1} \et2^{-1}\Gamb2^{-2} x_{i,-1} )^{1/2}$,
if we normalize to abundances $x_i\sim 10^{-1} x_{i,-1}$; there would be similar
resonant lines for other ion species, since hydrogenic ions have similar $f$ 
and $A_{ul}\lambda_{lu}^2$ values. While such widths would be significant, bulk 
velocity broadening (see below) would smear out any line features even more.  
Moreover, absorption lines would be partially compensated by emission 
from the blobs. 

It would be tempting to speculate that such features could be associated with
the lines reported by Ginga (e.g. Murakami et al., 1988, Fenimore, et al., 
1988). However, this would require special circumstances leading to a fairly 
narrow range of blob velocites, which might only be present in a small fraction 
of all cases. In general, any spectral features will be spread out due to the 
range of bulk Lorentz factors $\Gamma$ sampled by the line of sight. Emission 
line features associated with recombination will be further broadened because, 
even for a given $\Gamma$, there would be contributions with different Doppler 
blue-shifts from material with velocity making different angles with our line 
of sight. Even for a single value of $\Gamma_b$ this would introduce a 
broadening by $(\Delta \nu /\nu)_{ang}\sim 0.3-0.5$. The effect of this is to 
smear by this amount the red wing of any of the above spectral features.  
This smearing, however, would not be as important for the deep edges discussed 
above (equation [8]), which would be expected to survive.
The maximum blob Lorentz factor is $\eta$,  but there would be a 
spread below this maximum, given by values of $\Gamma_b$ for which $\calN$
exceeds the value (\ref{eq:calN}).
Slower blobs moving towards the observer take longer ($\propto \Gamma_b^{-2}$
in observer time) to reach a given radius. Therefore, early in the burst only
high- $\Gamma$ blobs will have reached the radius ($\sim c t_v \eta^2$) where
internal shocks occur. However, when the burst has been active for times $\gg 
t_v$, slower blobs whose Lorentz factor is of order $\eta (t_{ob}/t_v)^{-1/2}$ 
will have had time to reach the location of the emission, where $t_{ob}$ is the
observer frame time measured from the start of the burst. This leads to an
increasing spread of absorbing blob Lorentz factors
\beq
(\Delta \Gamma / \Gamma)_b \simeq
(\Gamma_{b,f}-\Gamma_b(t_{ob})/\Gamma_{b,f} =
1 - (t_{ob,o} / t_{ob} )^{1/2}~,
\label{eq:delgammab}
\enq
where $t_{ob,o} \simeq 10^{-1}\r13\Gamb2^{-2}$ s is the observer frame blob
dynamic time at $\r13 \sim 1$ (which in the wind regime used here is unrelated
to the burst duration). All lines, edges and maxiumum blanketing energies will 
therefore have an increasing spread 
$\Delta\nu/\nu \sim (\Delta\Gamma /\Gamma)_b$ with the time dependence of
equation (10), 
extending from an upper value 
corresponding to $\Gamma_{b,f}$ down to a lower limit which moves to softer 
energies in time.  The FeXXVI bound-free absorption will therefore move from 
blanketing the range $0.9\Gam2 - 2.2\Gam2$ MeV down to blanketing the range 
$0.09\Gam2 - 2.2 \Gam2$ MeV in a time $\sim 10 \r13\Gamb2^{-2}$ s after the 
burst starts.

\section {Conclusions}

Even though the emission from gamma-ray bursts is primarily non-thermal, we
have shown that the observed spectrum may be substantially modified by the
presence of highly ionized thermal plasma, with blueshifts of 10-100.  
The rate of absorption and re-emission by a thermal plasma, per unit mass, 
scales with density; the high ambient and ram pressure of the relativistic 
outflow  can confine plasma to such high densities that only a very small 
total mass can have conspicuous effects. The  material would be in a 'mist' 
of  blobs or filaments filling a small fraction of the volume, but which are 
individually so small that they provide a significant covering factor.  Even 
though small, these blobs can be envisaged as fluid-like because the gyroradius 
in megagauss magnetic fields is much smaller still. They can be accelerated 
to relativistic speed, without necessarily being disrupted, by the momentum 
of the jet-like outflow or by radiation pressure.  This material may be debris 
from a disrupted neutron star, e.g. \Mesz \& Rees, 1997b (in which case it could 
be highly enriched in heavy elements), or entrained from the boundaries of 
the jet in a 'hypernova' (e.g. Paczy\'nski, 1998) model. 

We obviously cannot predict how much material would be expelled in this form,
nor how the conditions near the central engine may evolve over the duration of
long bursts.  Nor do we know how the blobs would be distributed across the jet,
though entrained material would tend to be more prominent near the boundaries
(i.e. angles of order $\theta$ from the axis) rather than on the axis. However,
some general trends seem generic to this picture. 

The most prominent feature would be absorption above the photoionization edge
of FeXXVI, leading to a feature at this energy (i.e. 9.3 Kev multiplied by the
appropriate Doppler shift). In prolonged and complex bursts, it is likely that 
the primary emission comes from a series of internal shocks, at a distance 
$10^{13} - 10^{14}$ cm from the compact object. We would expect the feature to 
shift towards lower energies, becase later in the burst there would be time 
for lower-$\Gamma$ material to have reached the location of the reverse shock. 
(If different sub-bursts occur in shocks at different radii, then the absorption
effects should be more conspicuous in those close in, and this may introduce
a scatter about the general tendency for the cut-off to soften towards the
end of long bursts.  Spectra as observed by BATSE (most photons measured being 
in the range 50 - 500 KeV) would tend therefore to indicate, for objects with 
Fe-rich blobs, a spectral softening in time. Initially the burst would be 
classified as an HE (having a high energy component in the fourth LAD channel 
above 350 KeV), later to become an NHE (without significant emission above 350 
KeV), with departures due to the previous scatter,  e.g. as reported by 
Pendleton, et al. 1998. Also, when an average temporal evolution of many bursts 
is considered, it has been shown by Fenimore, 1998 that there is a clear
trend towards softening as the burst progresses. 
While there are alternative 
explanations for this softening, such as slowing down and cooling of the 
emitting material, we suggest that absorption of the kind discussed in this 
paper (characterized by the time dependence of equation [10]) may be relevant 
to such correlations.

\acknowledgements{This research has been supported by NASA NAG5-2857 and
the Royal Society}.

%\input{edgref.tex}
%\end{document}
%\newpage

\end{document}